# Dynamics, Structure and Glass Formation of Calcium Aluminate Liquids

Hao Liu[1,2], Ruikun Pan[1], Wenlin Chen[3], Zhitao Shan[1], Ang Qiao[1], Sandro Jahn[4], James W. E. Drewitt[5], Louis Hennet[6], David P. Langstaff[3], Haizheng Tao[1], G. Neville Greaves[1,3,7]*, Yuanzheng Yue[1,2]*

[1] *State Key Laboratory of Silicate Materials for Architectures, Wuhan University of Technology, Wuhan, 430070, People's Republic of China.*

[2] *Department of Chemistry and Bioscience, Aalborg University, DK-9220 Aalborg, Denmark.*

[3] *Department of Physics, Aberystwyth University, Penglais Campus, Aberystwyth, Ceredigion, SY23 3BZ, United Kingdom.*

[4] *GFZ German Research Centre for Geosciences, Telegrafenberg, 14473 Potsdam, Germany.*

[5] *School of Earth Sciences, University of Bristol, Wills Memorial Building, Queens Road, Clifton BS8 1RJ, Bristol, United Kingdom.*

[6] *Conditions Extrêmes et Matériaux: Haute Température et Irradiation, University d'Orléans, 1d avenue de la Rechearche Scientifique, 45071 Orléans cedex 2, France.*

[7] *Department of Materials Science and Metallurgy, University of Cambridge, 27 Charles Babbage Road, Cambridge CB3 0FS, United Kingdom.*

*Corresponding author. Email: gng25@cam.ac.uk (G.N.G.); yy@bio.aau.dk (Y.Z.Y.)

Crystalline calcium-aluminates include the phases essential in the setting of Portland cements developed over the last century. It is only within recent decades, however, that calcium-aluminate melts and glasses have begun to attract attention, bringing new functionalities in photonic and electronic applications. These studies, though, have been limited to compositions close to a deep eutectic from where glasses easily form. With the development of contactless levitation furnaces the glass forming region can now be hugely extended. We have taken advantage of these developments to rationalise, for the first time, melt rheology with structural properties across this expanded compositional range, substantiating this with atomistic simulation. In the process, we have discovered that supercooled calcium-aluminates comprise a new system where fragile-to-strong phase transitions are ubiquitous. Taking this holistic approach, we have quantified the common basis of thermo-physical and structural diversity in this novel glass-forming system, together with its inherent polyamorphism.

Calcium-aluminates (CaO-Al$_2$O$_3$), whose crystalline phases are critical agents in cement technology (*1*) and more recently in corrosion-resistant bioceramics (*2*), are also important components of the earth's mantle, with implications for magma-related processes (*3*). Latterly CaO-Al$_2$O$_3$ liquids and glasses have excited much attention, for example, with the development of infra-red windows, optical fibres and photometry materials (*4-6*) and the recent discovery of stable oxide electrides (*7,8*). More generally, however, it is the remarkable oxide glass-forming system on which they are based that is proving of fundamental importance. Compared to SiO$_2$, the classical glass-former familiar in oxide glasses (*9*), Al$_2$O$_3$ is atypical, deficient in oxygen, and generally defies Zachariasen's rules of corner-sharing tetrahedra required perpetuate a continuous random network (CRN) (*10*). Calcium-aluminates therefore demand fresh ways of understanding glass forming ability (GFA).

Originally interest centred on the mid composition (CaO)$_{12}$(Al$_2$O$_3$)$_7$, or C12A7, which coincides with a deep eutectic from which a stable glass can readily be formed using conventional melt-quenching techniques (*11*). With the development of laser-heated aerodynamic levitation furnaces (ALF) (*12-16*), however, glass compositions ranging from CA2 to C3A can now be readily prepared (*8,17*), and, with vapour deposition techniques, extending to amorphous Al$_2$O$_3$ (*18*) – representing 75% of CaO-Al$_2$O$_3$ compositions, greater than any other binary glass system (*9*). Moreover, structural properties of CaO-Al$_2$O$_3$ liquids can now also be accessed using contactless ALFs (*19-22*), with possibilities recently emerging to access thermo-physical properties (*15,16*). The structural source of the continuity of glass formation across most of the CaO-Al$_2$O$_3$ system appears to lie in the variety of cation and oxygen configurations seldom found in conventional oxide glasses (*17-20*). With increasing CaO content, configurational proportions modulate as liquids and their glasses evolve from compensated oxygen deficient random networks (ODRNs), through CRNs (*9,21*) to incomplete random networks (IRNs) (*22*).

The shear viscosity $\eta$ of most oxides, is approximately Arrhenian around the melting point $T_m$, but, towards the glass transition temperature $T_g$ it becomes less so, characterized by the fragility index *m*, defined in Angell plots by $m = \left(\frac{d(log\eta)}{d(T_g/T)}\right)_{T=T_g}$ (*23,24*). For typical *fragile* liquids *m* falls between 40-50. As supercooled temperatures approach ~$1.2T_g$ the ergodicity of the melt is lost (*25*), slow relaxation processes emerge (*9,24,26-28*) and the reciprocity between viscosity and ionic diffusion is broken, and a dynamic cross-over in structural heterogeneity occurs (*24-30*). In contrast, for liquids of archetypal glass-formers like $SiO_2$, the viscosity is Arrhenian throughout the supercooled region, with *m*~20 constituting *strong* behaviour. This fragile-strong demarcation in glass-forming liquids has been broken with the discovery of *fragile-to-strong* (*f-s*) viscosity transitions ~ $1.2T_g$, in glass-forming metals, water and network liquids (*31-38*).

To address these fundamental questions, we have measured, for the first time, thermo-physical properties across the extended $CaO-Al_2O_3$ glass-forming system, from supercooled to thermo-dynamically stable liquids. Using ALF techniques and large scale Molecular Dynamics (MD) simulations (fig. S1) we find close agreement (Fig.1). Differential Scanning Calorimetry (DSC) has revealed a minimum in $T_g$ for glasses coinciding with the eutectic in $T_m$ (Fig. 2), along with exceptionally low melt Coefficients of Thermal Expansion (CTEs) (fig. S2). By modelling Raman spectroscopy (fig. S3) and neutron scattering data of melts alongside viscosity $\eta$, the wide variety of Al(O) and O(Al) configurations common to atomic structure and rheology can be reconciled (Fig. 3). Finally, we have discovered that *f-s* transitions are ubiquitous in supercooled $CaO-Al_2O_3$ melts between low density and high density phases (LDL and HDL respectively), with volume and entropy changes at transition temperature $T_{f-s}$ peaking at the eutectic (Fig. 4).

Turning to Fig. 1, experimental log $\eta$ versus $T_m/T$ results across the calcium-aluminate glass-forming system rise from CA2 to C12A7 and decline to C5A. These are replicated in extensive 12,000 atom 50 ns MD simulations of diffusivities $D$ inverted to $\eta$ via the Eyring equation (*39*) (Fig. 1B), displaying a sharp maximum (Fig. 2D) coinciding with the deep eutectic at C12A7 (Fig. 2C). Melt CTEs obtained from ALF atomic density experiments (Fig. 2B and fig.S2) are exceptionally low at midway compositions, pointing to deep interatomic potentials, suggesting liquid CRN structures with improved atomic packing. DSC scans of CaO-Al$_2$O$_3$ glasses (Fig. 2A) demonstrate $T_g$ and heterogeneous crystallization temperatures ($T_p$) on heating quenched glasses both also follow the $T_m$ eutectic, with implications for varying GFA. Likewise, the other metrics of GFA – $T_g/T_m$, $\eta(T_m)$ and the crystal-melt density mismatch at $T_m$ $\Delta\rho/\rho_{melt}$ – clearly differentiate good (C12A7) from poor (CA2, C5A) glass formers (Fig. 2D). $T_g/T_m$ and $\eta(T_m)$ reflect how the depth of super-cooling reduces with improved GFA, being least for C12A7. Moreover, $\Delta\rho/\rho_{melt}$ follows the same profile, showing how heterogeneous crystallization close to $T_m$ (*40*) (avoided under contactless conditions (*16*)) preferentially favours nucleation of poor GFA phases like C3A ahead of good GFA phases like C12A7, important in Portland cements (*1*).

Next, the experimental atomic volume $V(T_m)$ is observed to rise monotonically as CaO replaces Al$_2$O$_3$, levelling-off at midway compositions (Fig. 2E). Modelling the structure of calcium aluminate melts and their glasses as two interpenetrating substructures – CaO and Al$_2$O$_3$ – there will be two percolation thresholds (fig. S4). Defined for a random network geometry as 0.2 by volume for (*41*), and by composition with CaO/(CaO+Al$_2$O$_3$) = 0.22 and 0.84, for CaO and Al$_2$O$_3$ respectively (Fig. 2E), these define the extensive glass-forming region, from ODRNs through CRNs to IRNs. Compared to $V(T_m)$ versus CaO/(CaO+Al$_2$O$_3$), the component atomic volumes, *viz.* $V[A](T_m)$ and $V[C](T_m)$, exhibit minima replacing the plateau in $V(T_m)$ indicating that C12A7 has the most well-packed network of the calcium aluminate melts and glasses, explaining its good GFA and low eutectic $T_m$. There are minima in the corresponding component free volumes at the eutectic, *viz.* $V_F[C](T_m)$ and $V_F[A](T_m)$ (fig. S4). Free volume generally assists ionic mobility while viscosity is dominated by the least mobile atoms (*39*), in this case O. Mid-range compositions, where network ions diffuse the least, contribute to the peak in $\eta(T_m)$ associated with the best GFA (Fig. 2E).

Moving on to the structural characteristics of CaO-Al$_2$O$_3$ glasses and melts, inter-atomic and inter-polyhedral units can be identified both experimentally and in MD simulations (Fig. 3). Raman spectra (Fig. 3A, B) reveal the expected rise in the boson peak (88 cm$^{-1}$) with decreasing atomic density (*9*), while the CaO rattling frequency (~115 cm$^{-1}$) falls. The Al-O-Al bond bending mode (~545 cm$^{-1}$) stays virtually constant – the main oxygen bridging angle being at ~120°, with the optic bond stretching Al-O frequency (~775 cm$^{-1}$) falling from IRN to CRN structures, but then increasing towards ODRN structures, pivoting around C12A7 with its efficient atomic packing.

Considering experimental neutron pair distribution functions $G(r)$ (*17,19,21,22*), taking average structures from our MD simulations of $\eta$ (Fig. 1B), predict $G(r)$ in close agreement with experiment (Fig. 3C), establishing a new and common basis connecting rheology with atomic structure in this unusual glass-forming system. Fig. 3E charts this in the envelope of Al and O configurations in melts as CaO replaces Al$_2$O$_3$ from atomistic simulations. Our discussion incorporates earlier observations on single compositions (*8,17,19,21,22*) but now with the advantage of considering the whole calcium-aluminate system, laid out in Fig. 3D for melts, with glasses exhibiting very similar proportions and trends. In particular, between the percolation thresholds as CaO/(CaO+Al$_2$O$_3$) rises, significant fractions of pentagonal Al$^V$ and O$^3$ triclusters are present in ODRN compositions, like CA2, similar to the same configurations reported in liquid Al$_2$O$_3$ (*19*). They also tend to be preferred neighbours in C12A7 (fig. S5) at which point structures approximate to CRNs. Thereafter both these configurations, foreign to conventional tetrahedral networks, decline in number, virtually disappearing by C3A, at which time non-bridging oxygens ($O^1$) exceed bridging oxygens ($O^2$); $Ca^{2+}$ over charge-compensating tetrahedral $Al^{3+}$, leads to a breakup of the corner-sharing $Al_2O_3$ network and the formation of IRNs. The diversity of configurations ensures a smooth progression, visualized in polyhedral images (Fig. 3D and fig. S6). Noticeably, isolated oxygens $O^0$ also appear by CA increasing in number to C5A, suggesting glasses, suitably reduced $Al_2O_3$, might solvate electrons, offering a wider range of electride melts and glasses (*7*).

Throughout the $CaO-Al_2O_3$ glass-forming system, calcium occupies distorted octahedral sites whose connectivity affects the density of melts and glasses (fig. S7). As $CaO/(CaO+Al_2O_3)$ increases in melts, isolated and small clusters are replaced initially by medium-sized clusters and ultimately by one large cluster beyond the $Al_2O_3$ percolation threshold (fig. S7). In glasses, this sequence of developing medium-range order starts at lower CaO content (fig. S7), suggesting an ordered phase as $T_g$ is approached.

Finally, and remarkably for all calcium-aluminate compositions examined, there is a mismatch between $\eta$ from ALF experiments close to $T_m$ and from DSC close to $T_g$, defining a dynamic fragile-strong crossover at $T_{f\text{-}s}$ (Fig. 4A) ~ 1.1-1.2 $T_g$ (Fig. 2C). Based on melt and glass densities, and CTE data (fig. S2), *f-s* transitions are associated with step-wise reductions at $T_{f\text{-}s}$ in atomic volume $\Delta V_{f\text{-}s}$ (Fig. 4C), substantiating LDL-HDL transitions. Analysis of $\eta$ ~$T_m$ and ~$T_g$ confirm complementary reductions in excess entropy $\Delta S_{f\text{-}s}$ (Fig. 4C), that also circumvent the Kauzmann paradox (*42*). In particular, the Claperyon slope $\frac{dT}{dP} = \frac{\Delta V_{f-s}}{\Delta S_{f-s}}$ is positive for super-cooled calcium-aluminates in the same sense as conventional melting, akin to metallic glass melts (*32-34*), but unlike the negative slopes of liquid-liquid transitions in network structures (*8,31,36*). While *dT/dP*>0 is usually attributed to non-directionally bonded icosahedral packed structures, calcium-aluminate liquids clearly have network topology, but can be envisaged as the close-packing of polyhedra (Fig. 3). Furthermore, the steepness follows GFA (Fig. 2D and Fig. 4C). Since contactless conditions were not possible around $T_{f\text{-}s}$, the expected reversibility of *f-s* transitions remains unexplored.

In the context of the Potential Energy Landscape *f-s* LDL-HDL transitions in calcium-aluminates involve transfer between fragile narrow shallow minima and strong broad deep minima, respectively, separated by $T_{f\text{-}s}\Delta S_{f\text{-}s}$, with even deeper minima for crystallization, both transitions releasing energy (*9*). This is illustrated schematically in fig. S10, distinguishing the solidification of melts with good GFA and steep *dT/dP* from those without.

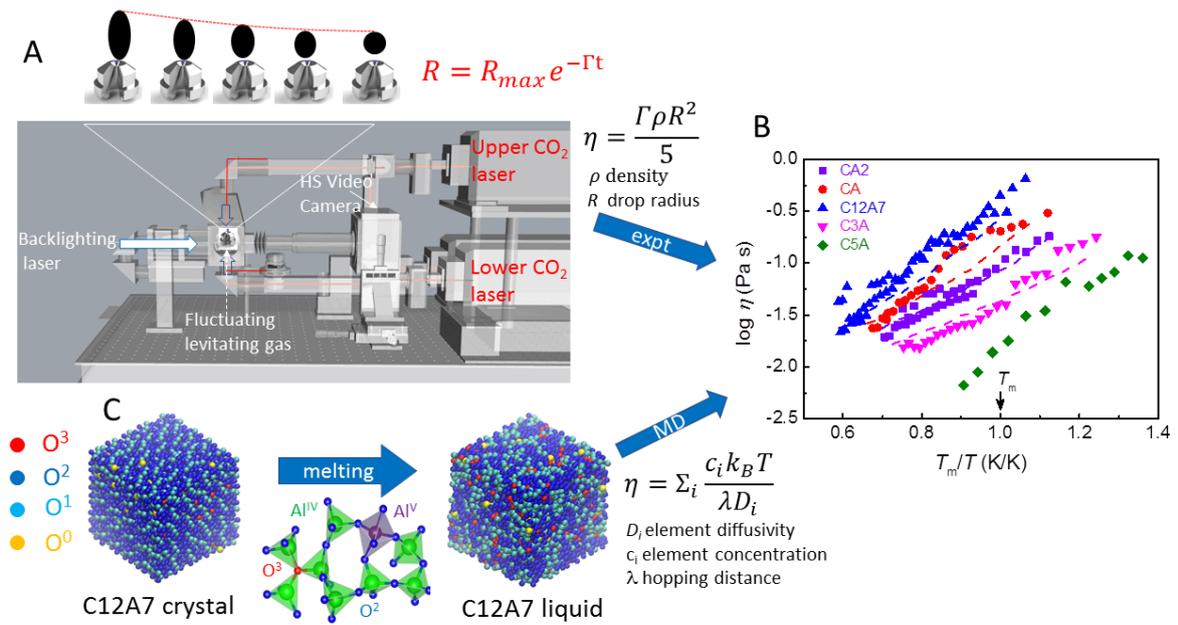

**Fig. 1. Viscosity measurement and modelling: (A)** AFL layout. **(B)** $\log \eta$ vs. $T_m/T$. 12,000 atom 50 ns MD simulations of $\eta$ (dashed lines) derived from diffusion (*38*), reproducing the magnitudes of $\eta$ and coefficients of thermal expansion (CTE) (fig. S1). **(C)** Visualization of distribution in atomic configurations around O occurring on melting.

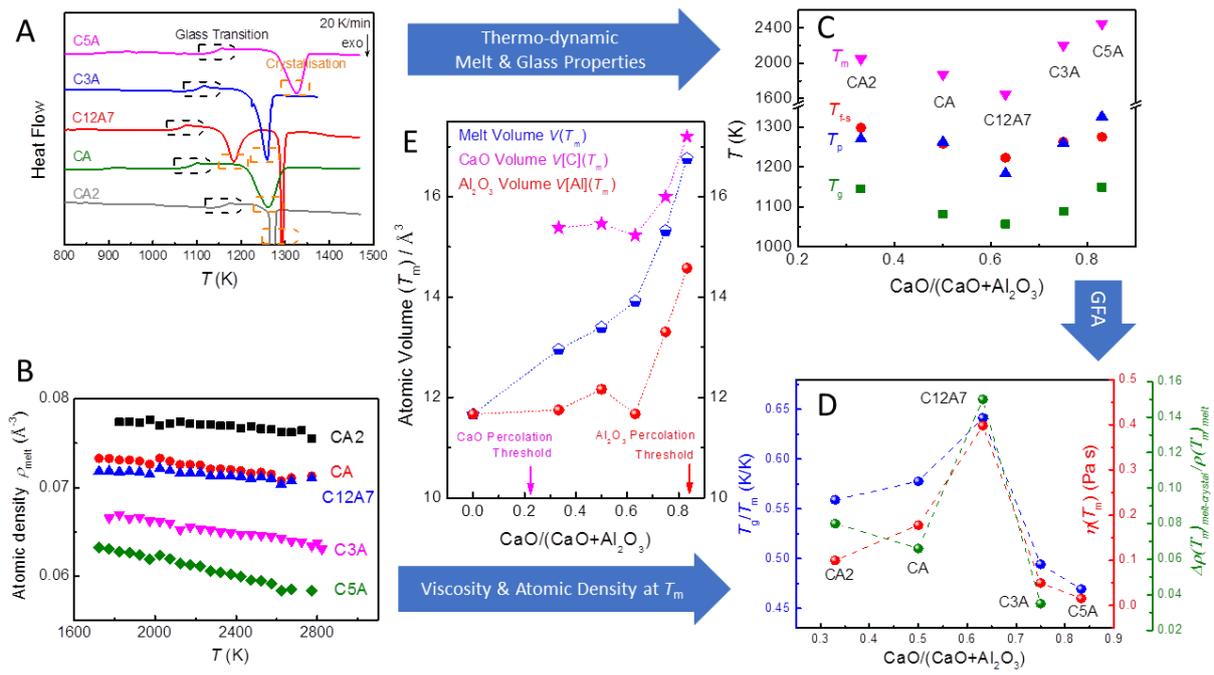

**Fig. 2. Thermo-physical Properties, GFA and Atomic Volume:** (**A**) DSC identifying $T_g$ and heterogenous crystallization temperature $T_p$ for glasses. (**B**) Melt density $\rho_{melt}$ from AFL experiments. (**C**) Profile of $T_g$, $T_p$, $T_{f-s}$ and $T_m$ sharing common minimum at C12A7 eutectic. (**D**) GFA metrics $T_g/T_m$, $\eta(T_m)$, $\Delta\rho(T_m)_{melt-crystal}/\rho(T_m)_{melt}$. (**E**) Atomic volume $V(T_m)$ from (B) and $Al_2O_3$ $V[A](T_m)$ and CaO $V[C](T_m)$ subcomponents.

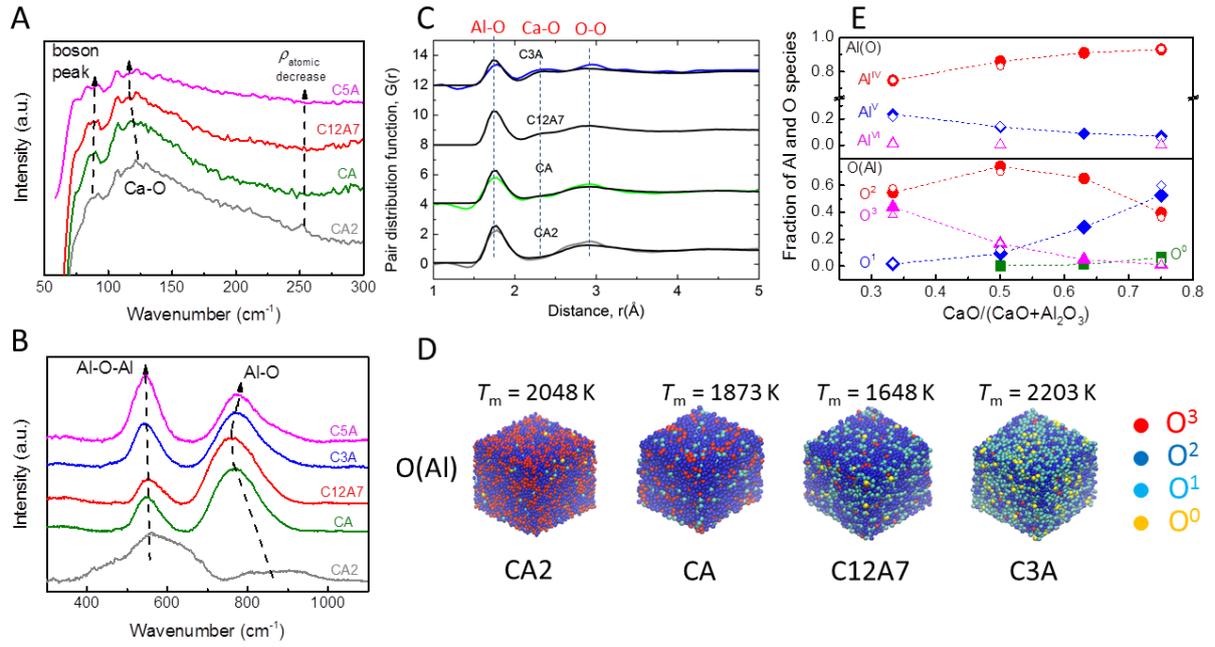

**Fig. 3. Structure, Dynamics, and Ion Configurations:** **(A)** Low frequency and **(B)** High frequency Raman spectra, showing compositional trends in the boson peak, Ca-O, Al-O-Al and Al-O modes. **(C)** Neutron scattering pair distribution functions $G(r)$ of melts (coloured) averaged MD simulations from $\eta$ modelling (grey). **(D)** Visualisation of O configurations from MD simulations. **(E)** Trends in fractions of Al and O configurations with CaO-Al$_2$O$_3$ composition; $\eta$-derived MD (open symbols) and $G(r)$-derived (closed symbols).

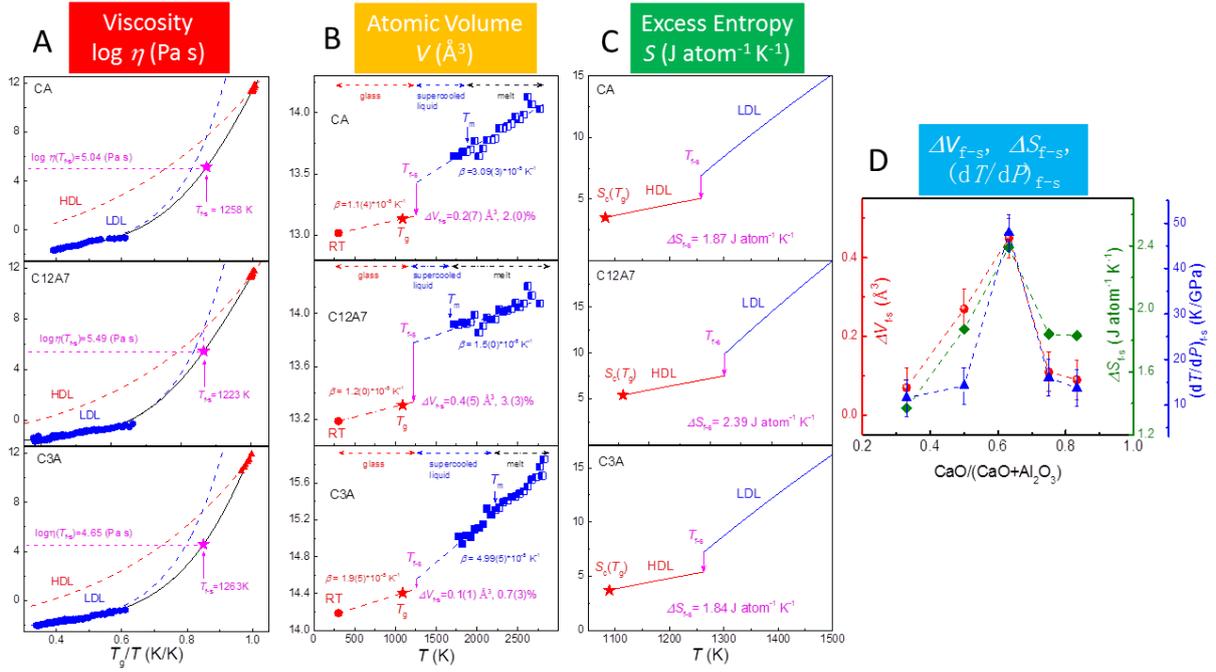

**Fig. 4. Fragile-to-Strong Transitions:** (A) *f-s* transitions identified from mismatch between melts around $T_m$ and $T_g$ analysed using the MYEGA equation identifying $T_{f\text{-}s}$, $\eta(T_{f\text{-}s})$ and LDL and HDL liquid phases for CA, C12A7 and C3A (*24,32*). (B) Volume steps $\Delta V_{f\text{-}s}$, (C) Entropy steps $\Delta S_{f\text{-}s}$ at $T_{f\text{-}s}$ for CA, C12A7 and C3A melts (SI). (D) $\Delta V_{f\text{-}s}$, $\Delta S_{f\text{-}s}$ and $dT/dP$ for calcium aluminate system.

**ACKNOWLEDGEMENTS**

We are grateful to the National Natural Science Foundations of China (no. 51372180) and the funding of State Key Laboratory of Silicate Materials for Architectures (SYSJJ2016-14), including the post of Strategic Scientist (GNG). We also thank J. C. Mauro for valuable discussions and Lars R. Jensen for the help with Raman measurements.

**SUPPLEMENTARY MATERIALS**

Materials and Methods

Supplementary Text

Figs. S1-S11